\documentclass[runningheads]{llncs}
\usepackage[T1]{fontenc}
\usepackage{graphicx}
\usepackage{amsmath}
\usepackage{amsfonts}
\usepackage{graphicx}
\usepackage{tikz}
\usepackage{mathtools}
\usepackage{hyperref}
\newcommand{\defeq}{\vcentcolon=}
\usetikzlibrary{arrows.meta,positioning,calc}
\usepackage{array, boldline, makecell, booktabs}

\begin{document}
\title{Longest-chain Attacks: Difficulty Adjustment and Timestamp Verifiability}
%
%
\author{Tzuo Hann Law\inst{1} \and
Selman Erol\inst{2} \and
Lewis Tseng\inst{3}}
\authorrunning{Law et al.}
%
\institute{Unaffiliated, 
\email{tzuohann@gmail.com} \and
CMU, 
\email{erol@cmu.edu} \and
Clark University, 
\email{lewistseng@acm.org}}
\maketitle              
\begin{abstract}
We study an adversary who attacks a Proof-of-Work (POW) blockchain by selfishly constructing an alternative longest chain. We characterize optimal strategies employed by the adversary when a difficulty adjustment rule al\`a Bitcoin applies. As time (namely the timestamp specified in each block) in most permissionless POW blockchains is somewhat subjective, we focus on two extreme scenarios - when time is completely verifiable, and when it is completely unverifiable. We conclude that an adversary who faces a difficulty adjustment rule will find a longest-chain attack very challenging when timestamps are verifiable.  POW blockchains with frequent difficulty adjustments relative to time reporting flexibility will be substantially more vulnerable to longest-chain attacks. Our main fining provides guidance on the design of difficulty adjustment rules and demonstrates the importance of timestamp verifiability.

\keywords{Bitcoin \and blockchain \and longest chain attack \and difficulty adjustment \and cryptocurrency \and Proof-of-Work}
\end{abstract}
\section{Introduction}
Permissionless Proof-of-Work (POW) consensus systems feature a network of peer-to-peer nodes that add blocks containing information to a blockchain. Since every node prefers blocks containing information specific to their own benefit, there will be no consensus unless a mechanism that grants nodes permission to add blocks is in place. Nakamoto \cite{Nakamoto2009} consensus is one such mechanism. Nodes earn the privilege to propose a block by reporting a suitable nonce to accompany a block containing specific information of their own choice. Receiving nodes accept the proposed block if it is valid. A nonce is a number embedded in the block so that the output of some cryptographic hash function of the block fulfills some condition. Typically, this condition takes the form of a difficulty threshold. An acceptable nonce is one which yields a block hash that falls within a specified distance from zero. Additionally, the information within the blocks must conform to a set of rules referred to as the blockchain's protocol. 

Due to the way that nonces are used, suitable ones can only be found by trial and error. The process of nonce finding is a race of who can test potential numbers most quickly and report their finding to the rest of the network. Nonce finding is literally work in the sense that the only practical way to do it is to pass electrical current through computer chips. A suitable nonce is quite simply, proof that work was done. Other nodes who receive this information then check if the newly received information contains the more \textit{cumulative} and \textit{valid} work than their own leading block as measured by the same metric.\footnote{Usually, the longest chain in a blockchain also requires the most amount of cumulative work to create. However, this is not always the case. For example, see Ethereum prior to its transition to Proof-of-Stake.} If it is the case, the receiving node would accept the newly minted block and build on top of it. This process is referred to as \textit{mining} and amounts to an arms race. Miners who control more (powerful) computer hardware and have access to cheap energy are able to test potential nonces more quickly. As a result, these miners add more blocks in their own favor. 

Without additional safeguards, such a system would imply that an increase in the system mining capacity would result in a higher rate of token generation. To stabilize the token generation rate, blockchain protocols typically specify a \textit{difficulty adjustment} protocol or rule. The difficulty is adjusted so that the token generation rate is steered towards some target rate as defined in the protocol. There are many different difficulty adjustment rules being used with this same overarching mandate. In this paper, we consider one that is modeled after the protocol used in Bitcoin \cite{Nakamoto2009}.

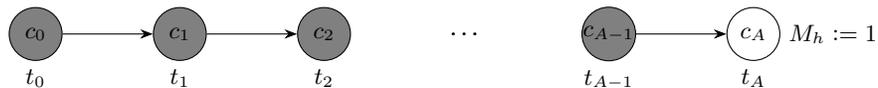
\begin{figure*}[!ht]
\begin{tikzpicture}[
      mycircle/.style={
         circle,
         draw=black,
         fill=gray,
         fill opacity = 0.3,
         text opacity=1,
         inner sep=0pt,
         minimum size=20pt,
         font=\small},
      mycircle2/.style={
         circle,
         draw=black,
         fill=white,
         fill opacity = 0.3,
         text opacity=1,
         inner sep=0pt,
         minimum size=20pt,
         font=\small},
      myarrow/.style={-Stealth},
      myarrow2/.style={-{Implies},double},
      node distance=0.6cm and 1.2cm
      ]
      \node[label =below:{$t_0$},mycircle] at (-5, 0) (c0) {$c_0$};
      \node[label =below:{$t_1$},mycircle,right=of c0] (c1) {$c_{1}$};
      \node[label =below:{$t_2$},mycircle,right=of c1] (c2) {$c_{2}$};
      \node[right=of c2] (dots1) {\ldots};
      \node[label =below:{$t_{A-1}$},mycircle,right =of dots1] (cAm) {$c_{A-1}$};
      \node[label =right:{$M_h \defeq 1$},label =below:{$t_A$},mycircle2,right =of cAm] (cA) {$c_A$};
      
    \foreach \i/\j/\txt/\p in {
      c0/c1//below,
      c1/c2//below,
      cAm/cA//below}
       \draw [myarrow] (\i) -- node[sloped,font=\small,\p] {\txt} (\j);

   
   \node at ($(c2)!.5!(cAm)$) {\ldots };
    \end{tikzpicture}
\caption{Simplified POW blockchain. The block identifier, subscripted by its blockheight (e.g. $c_i,c_A$) refers to the block as well as the unique blockchain formed from tracing the block back to $c_0$. $t_{i}$ is the reported timestamp for block $c_i$. The frame of reference is $t_A$ where $c_0$ to $c_{A-1}$ are mined (in grey) and $c_A$ (in white) has been constructed, but not mined (unknown suitable nonce).}
\label{fig:honestonly}
\end{figure*}
\vspace{-20pt}

Ideally, a difficulty adjustment rule would adjust the difficulty level according to the mining capacity of the network since that is the primary determinant of the block-finding rate. However, the mining capacity of the network is unobservable and particularly so in permissionless systems. Instead, the difficulty adjustment algorithms utilize the time taken for successive blocks to be created as an estimate of mining capacity and this is in turn ``\textbf{proxied}'' by the timestamps reported in each block. We emphasize the word ``proxied'' because the system time of a different node is itself unobservable due to the nature of asynchronous networks. Nodes can report any timestamp they please so long as the reported timestamp conforms to some protocol which in turn ensures its acceptance by other nodes. In addition, nodes can successfully mine a block and not report their success until a later time. Since there are various protocols for accepting/rejecting timestamps, there is substantial variation in the flexibility nodes have for timestamp reporting across different POW blockchains. 

We investigate how this timestamp flexibility in relation to the difficulty adjustment rule influences the \textit{optimal strategy} that an adversary employs when mounting a longest-chain attack. To make key ideas clearer and more transparent, we work with deterministic mining, which was also adopted in \cite{BitcoinDDoS_FC14,coin-hopping_ESORICS17}. 
The strategies that apply in a deterministic setting continue to apply in the appropriate probabilistic setting under the right assumptions about the adversary's preferences. Other than benign analytical oddities such as integer constraints, none of our key points and findings depends on the deterministic mining assumption that we make.

In this paper, we characterize the optimal strategies in a simplified POW blockchain with deterministic mining \cite{BitcoinDDoS_FC14,coin-hopping_ESORICS17} and comment on how the insights from our analysis would translate to real-world implications. Our main finding is that difficulty adjustment rules offer substantial protection against longest-chain attacks provided timestamps are accurate relative to the frequency of the difficulty adjustment.

\section{Related Work}
We discuss the closely related works that investigate the effect of varying mining difficulty and attacks on manipulating timestamps or mining difficulty. Garay et al. \cite{Garay_BitcoinBackbone_Eurocrypt15} formalize and analyze the core of the Bitcoin protocol, namely the Bitcoin
backbone, in the static setting (with fixed number of nodes and fixed difficulty). Subsequently, Garay et al. \cite{Garay_BitcoinBackbone_varyDifficulty_Crypto17} extend and
analyze Bitcoin backbone protocol with  mining difficulty adjustment by formulating the target (re)calculation function in Bitcoin. Kraft \cite{Kraft_BitcoinDifficulty_P2P16} and Noda et al. \cite{Noda_BitcoinDifficultyEcon_EC20} study the effect of mining difficulty adjustment and block arrival rate.

The notion of selfish mining is first proposed in \cite{Eyal_SelfishMining_CACM18}, which demonstrates disobedient mining could be more profitable than being honest, i.e., following the Bitcoin specification. Subsequently, Davidson and Diamond  \cite{Davidson_SelfishDifficulty_IACR20}  and Alarc{\'{o}}n Negy et al. \cite{GunSirer_SelfishDifficulty_FC20} investigate the profitability of selfish mining \cite{Eyal_SelfishMining_CACM18} under different difficulty
adjustment mechanism using simulation. In particular, it is shown that an intermittent selfish mining strategy \cite{GunSirer_SelfishDifficulty_FC20} gives the attacker a higher profit by performing selfish mining intermittently. Per time-unit profitability of selfish mining under different difficulty adjustment mechanisms is also identified in  \cite{GunSirer_SelfishDifficulty_FC20}. 

Meshkove et al. \cite{coin-hopping_ESORICS17} introduce the coin-hopping attack which allows attackers to gain profit compared to honest players by exploiting the mining difficulty adjustment. Boverman  \cite{Timejacking} describes an attack which forces honest players to accept a chain other than the canonical chain by tampering the ``network time counter'' at an honest player or even a majority of
players. Fiat et al. \cite{Fiat_EnergyAttack_EC19} and Goren and Spiegelman \cite{Goren_ElectricityAttack_EC19} introduce the Energy Equilibria attack which minimizes operational (energy) costs and manipulate mining difficulty to increase mining rewards per unit of time on average. 

Yaish et al. \cite{Yaish_StretchSqueeze_EC22} introduce an approach, called stretching and squeezing, which can create
and exploit interest-rate arbitrage between decentralized finance platforms by manipulating the mining difficulty. They also find two timestamp weaknesses in Geth’s code.  Later, Yaish et al. \cite{Yaish_Uncle_IACR22} identify an attack, called uncle maker, that allows attacker to gain higher profit by manipulating block timestamps at proper times. 

To the best of our knowledge, we are the first to investigate  the \textit{joint effect} of difficulty adjustment and timestamp verifiability on the optimal strategy for mounting longest-chain attacks.

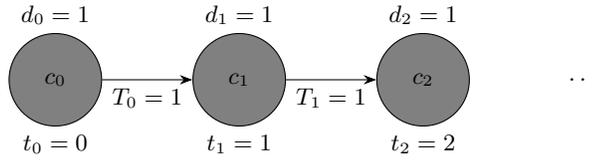
\begin{figure}[t]
\begin{tikzpicture}[
      mycircle/.style={
         circle,
         draw=black,
         fill=gray,
         fill opacity = 0.3,
         text opacity=1,
         inner sep=0pt,
         minimum size=35pt,
         font=\small},
      mycircle2/.style={
         circle,
         draw=black,
         fill=white,
         fill opacity = 0.3,
         text opacity=1,
         inner sep=0pt,
         minimum size=35pt,
         font=\small},
      myarrow/.style={-Stealth},
      myarrow2/.style={-{Implies},double},
      node distance=0.6cm and 1.2cm
      ]
      \node[label =below:{$t_0 = 0$},label =above:{$d_0 = 1$},mycircle] at (-5, 0) (c0) {$c_0$};
      \node[label =below:{$t_1 = 1$},label =above:{$d_1 = 1$},mycircle,right=of c0] (c1) {$c_{1}$};
      \node[label =below:{$t_2 = 2$},label =above:{$d_2 = 1$},mycircle,right=of c1] (c2) {$c_{2}$};
      \node[right=of c2] (dots1) {\ldots};
      
    \foreach \i/\j/\txt/\p in {
      c0/c1/$T_0 = 1$/below,
      c1/c2/$T_1 = 1$/below}
       \draw [myarrow] (\i) -- node[sloped,font=\small,\p] {\txt} (\j);

    \end{tikzpicture}
\caption{With the assumptions made, the canonical chain grows constantly a block per unit of time.}
\label{fig:boringtime}
\end{figure}

\section{\textbf{Simplified Proof-of-Work Blockchain}}

\paragraph{Honest miners.}
We model a vastly simplified Proof-of-Work (POW) blockchain with genesis block $c_0$ as shown in Figure \ref{fig:honestonly}. Honest miners control $M_h\defeq 1$ mining capacity and do \textit{not} behave strategically. They naively extend the longest chain known and ignore all others. In our notation, block $c_i$ is defined, but yet to be mined at time $t_i$, i.e., its header is fixed but a suitable nonce remains unknown. The only active miners prior to $t_A$ are the honest miners. At that time, honest miners have mined $c_0$ to $c_{A-1}$, shaded grey, and are about to start mining block $c_A$. An adversary which we will later describe initiates an attack at time $t_A$.

\begin{figure*}[t]
\begin{tikzpicture}[
      mycircle/.style={
         circle,
         draw=black,
         fill=gray,
         fill opacity = 0.3,
         text opacity=1,
         inner sep=0pt,
         minimum size=20pt,
         font=\small},
      mycircle2/.style={
         circle,
         draw=black,
         fill=white,
         fill opacity = 0.3,
         text opacity=1,
         inner sep=0pt,
         minimum size=20pt,
         font=\small},
      myarrow/.style={-Stealth},
      myarrow2/.style={-{Implies},double},
      node distance=0.6cm and 1.2cm
      ]
      \node[label =below:{$t_0$},mycircle] at (-5, 0) (c0) {$c_0$};
      \node[label =below:{$t_1$},mycircle,right=of c0] (c1) {$c_{1}$};
      \node[right=of c1] (dots1) {\ldots};
      \node[label =below:{$t_A$},mycircle,right =of dots1] (cA) {$c_A$};
      \node[right=of cA] (dots2) {\ldots};
      \node[label =right:{$M_h \defeq 1$},label =below:{$t_N$},mycircle2,right =of dots2] (cN) {$c_N$};
      \node[mycircle2,below =of c1] (a1) {$a_{1}$};
      \node[mycircle2,below =of cN] (aN) {$a_{N}$};
    \node[label =right:{$M_a > 1$},label =below:{$\tilde{t}_{a_{N+1}} \leq t_{N}$},mycircle2,right =of aN] (aNp) {$a_{N+1}$};

    \foreach \i/\j/\txt/\p in {
      c0/c1//below}
       \draw [myarrow] (\i) -- node[sloped,font=\small,\p] {\txt} (\j);

    \foreach \i/\j/\txt/\p in {
      aN/aNp//below,
      c0/a1//below}
       \draw [dashed,->] (\i) -- node[sloped,font=\small,\p] {\txt} (\j);

   \node at ($(a1)!.5!(aN)$) {\ldots };
\end{tikzpicture}
\caption{Attack starts at time $t_A$. Adversary must construct some chain $a_{N+1}$ that is at least a block longer than the honest miner's chain. In this figure, we align both chains by block height. Note that $\tilde{t}_{N+1}$'s timestamp must be no more than $t_N$.}
\label{fig:withattack}
\end{figure*}
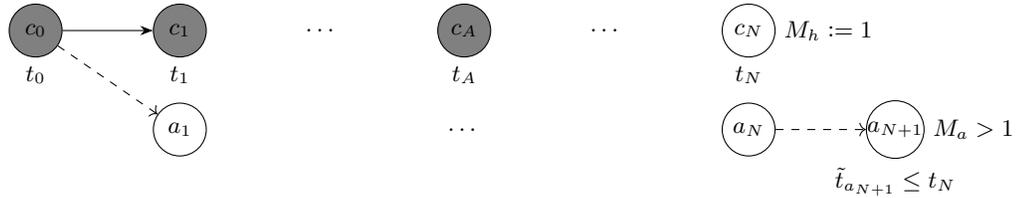


\paragraph{Deterministic Mining.}
With deterministic mining \cite{BitcoinDDoS_FC14,coin-hopping_ESORICS17}, if $M_i$ mining power is dedicated to a block with difficulty $d_i$, then a suitable nonce will take  
\begin{equation} \label{blockfindtime}
	T_i = \frac{d_i}{M_i}
\end{equation}
time units to be discovered. 

The target block finding rate is 1 block per-unit time and each difficulty epoch contains one block. In the spirit of Bitcoin's difficulty adjustment rule, the difficulty adjusts with every block/epoch following 
\begin{equation} \label{diffadj}
\begin{gathered}
	d_{i+1}  = \frac{d_i}{\tilde{T}_{i}} = \frac{d_i}{(\tilde{t}_{i+1} - \tilde{t}_{i})}\\
    d_0 = 1    
\end{gathered}
\end{equation} where $\tilde{t}$ refers to the timestamps reported in the blocks. 

Combining Equations (\ref{blockfindtime}) and (\ref{diffadj}) yields 
\begin{equation*}
	d_{i+1} = \frac{d_i}{\tilde{T}_{i}} = \frac{d_{i-1}}{\tilde{T}_{i-1}\tilde{T}_{i}} = \frac{d_{0}}{\prod_{j=0}^i\tilde{T}_{i-j}}
\end{equation*} 

As in \cite{Nakamoto2009,coin-hopping_ESORICS17}, we assume no propagation latency effects and no block transition delays. Honest miners propagate a block as soon as its nonce is discovered and immediately start mining on the next block. On the honest chain, \begin{equation*}
	d_{i+1} = \frac{d_i}{T_{i}} = M_{i}
\end{equation*} 
and consequently, in our setup, $d_i = T_i = M_h = 1$ and that $t_i = i$ as shown in Figure \ref{fig:boringtime}.

\paragraph{Adversary.}
Our adversary seeks to replace $c$ with a longer chain without any concern for the value of $c$ or for the energy costs consumed. More concretely, the adversary could be a hostile government or sophisticated hacker, a large short financial position on $c$'s tokens, or a competing blockchain. Given recent developments in quantum computing, it would not be a stretch to think of the adversary as a miner armed with quantum computing capabilities as described in Bard et al.  \cite{Bard2021QuantumAO}.

The adversary initiates the longest-chain attack at time $t_A$ which coincides with the time honest miners start mining block $c_A$. The adversary controls mining capacity $M_a > M_h$ and seeks to replace $c$ by constructing an alternative chain starting at $c_0$. This analysis also applies if the adversary attacks a blockchain $A$ blocks behind its most recent block. We name this adversarial chain $a$. We use $\tilde{t}$ to denote timestamps as reported by the adversary for chain $a$.

While honest miners are extending $c_A$, the adversary must construct $a_{1}$, $ a_{2}$, ..., $a_{N+1}$ where $a_{N+1}$ is the first block higher than $c_N$ which is the terminal block on the chain $c$. Revealing $a_{N+1}$ to the honest miners ends the game in favor of the adversary and invalidates all transactions between $c_N$ and $c_{1}$ due to the longest-chain rule. The adversary also seeks to end the game as \textit{quickly} as possible as measured by the time it spends on the attack. Since chain $a$ is revealed at $t_N$, $\tilde{t}_{a_{N+1}}$ must be less than or equal to $t_N$. The timestamp associated with $a_1$ is $t_1$ because $c_1$ and $a_1$ share a common ancestor and we assume no network propagation delay. Finally, we assume in this paper that the adversary does not dedicate any of its mining capacity to the canonical chain and operates stealthily until it reveals chain $a$.

With all that we have described, what is the best that the adversary can do? As was suggested earlier in this paper, the answer will depend on how much flexibility the adversary has in reporting timestamps which is unobservable by honest miners.

\section{\textbf{What is the Time?}}
Time is a subjectively defined object in Bitcoin Core's protocol. As of January 2023, a node accepts a block if the timestamp reported is within lower and upper bounds.\footnote{Block timestamp \url{https://en.bitcoin.it/wiki/Block_timestamp}} The lower bound is the median timestamp of the node's previous 11 blocks. The upper bound is the median time reported by other connected nodes plus two hours. In other words, ``time'' specified in a block is valid so long it is not from the distant past or future. 

This rather fluid notion of time directly affects the difficulty adjustment rule. Specifically, difficulty adjustment ensures that fluctuations in mining capacity changes will result in temporary deviations from the target token production schedule. Since this rule applies to all chains, an adversary who chooses to mount a longest-chain attack must also adhere to this rule while constructing chain $a$. In order to determine how the difficulty adjustment rule affects the adversary's attack, we must now be precise about the nature of time.

\begin{figure*}[t]
\begin{tikzpicture}[
      mycircle/.style={
         circle,
         draw=black,
         fill=gray,
         fill opacity = 0.3,
         text opacity=1,
         inner sep=0pt,
         minimum size=20pt,
         font=\small},
      mycircle2/.style={
         circle,
         draw=black,
         fill=white,
         fill opacity = 0.3,
         text opacity=1,
         inner sep=0pt,
         minimum size=20pt,
         font=\small},
      myarrow/.style={-Stealth},
      myarrow2/.style={-{Implies},double},
      node distance=0.6cm and 1.2cm
      ]
      \node[label =below:{$t_0$},mycircle] at (-5, 0) (c0) {$c_0$};
      \node[label =below:{$t_1$},mycircle,right=of c0] (c1) {$c_{1}$};
      \node[right=of c1] (dots1) {\ldots};
      \node[label =below:{$t_A$},mycircle2,right =of dots1] (cA) {$c_A$};
      \node[label =below:{$t_{A+1}$},mycircle2,right =of cA] (cAp) {$c_{A+1}$};
      \node[right=of cAp] (dots1) {\ldots};
      \node[right=of dots1,label =below:{$t_{N}$},mycircle2] (cN) {$c_N$};
      \node[label =below:{$\tilde{t}_{a_1} = t_{1}$},mycircle2,below =of c1] (a1) {$a_{1}$};
      \node[label =below:{$\tilde{t}_{a_2} = t_A + \frac{d_{a_1}}{M_{a_1}}$},mycircle2,right = 4cm of a1] (a2) {$a_{2}$};
      \node[label =below:{$\tilde{t}_{a_3} = \tilde{t}_{a_2} + \frac{d_{a_2}}{M_{a_2}}$},mycircle2,right = 1.6cm of a2] (a3) {$a_{3}$};
      \node[label =below:{$\tilde{t}_{a_{N+1}} \leq t_N$},mycircle2,below = of cN] (aNp) {$a_{N+1}$};
      
    \foreach \i/\j/\txt/\p in {
      c0/c1//below,
      c0/a1//below}
       \draw [myarrow] (\i) -- node[sloped,font=\small,\p] {\txt} (\j);

    \foreach \i/\j/\txt/\p in {
      cA/cAp//below,
      a1/a2//below,
      a2/a3//below}
       \draw [dashed,->] (\i) -- node[sloped,font=\small,\p] {\txt} (\j);

\node at ($(a3)!.5!(aNp)$) {\ldots };
\end{tikzpicture}
\caption{We now align both chains by time. The adversary reports all timestamps truthfully (because it has to). Notice that $a_2$ is created after $t_A$ because the adversary started mining at time $t_A$. $a_{N+1}$ must be timestamped no later than $t_N$.}
\label{fig:veriftime}
\end{figure*}
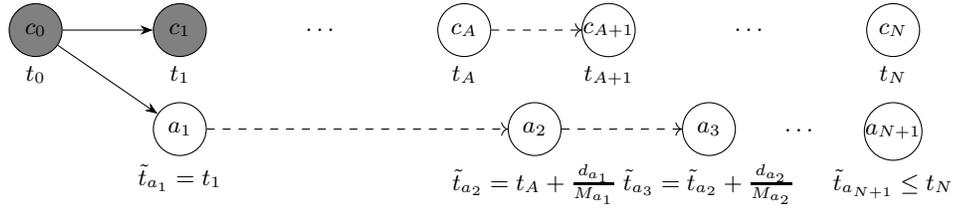

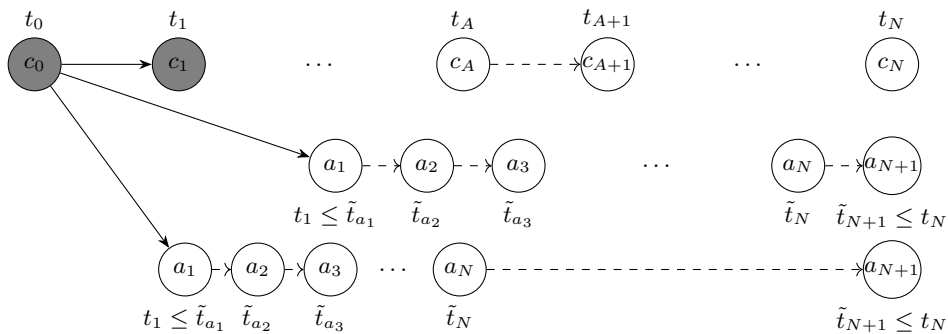
\begin{figure*}[t]
\begin{tikzpicture}[
      mycircle/.style={
         circle,
         draw=black,
         fill=gray,
         fill opacity = 0.3,
         text opacity=1,
         inner sep=0pt,
         minimum size=20pt,
         font=\small},
      mycircle2/.style={
         circle,
         draw=black,
         fill=white,
         fill opacity = 0.3,
         text opacity=1,
         inner sep=0pt,
         minimum size=20pt,
         font=\small},
      myarrow/.style={-Stealth},
      myarrow2/.style={-{Implies},double},
      node distance=0.6cm and 1.2cm
      ]
      \node[label =above:{$t_{0}$},mycircle] at (-5, 0) (c0) {$c_0$};
      
      \node[label =above:{$t_{1}$},mycircle,right=of c0] (c1) {$c_{1}$};
      \node[right=of c1] (dots1) {\ldots};
      \node[label =above:{$t_{A}$},mycircle2,right =of dots1] (cA) {$c_A$};
      \node[label =above:{$t_{A+1}$},mycircle2,right =of cA] (cAp) {$c_{A+1}$};
      \node[right=of cAp] (dots1) {\ldots};
      \node[right=of dots1,label =above:{$t_{N}$},mycircle2] (cN) {$c_N$};

      \node[label =below:{$\tilde{t}_{N+1}  \leq t_N$},mycircle2,below = of cN] (aNp) {$a_{N+1}$};
      \node[label =below:{$\tilde{t}_{N}$},mycircle2,left = 0.5cm of aNp] (aN) {$a_{N}$};
      \node[label =below:{$\tilde{t}_{a_3}$},mycircle2,left = 3cm of aN] (a3) {$a_{3}$};
      \node[label =below:{$\tilde{t}_{a_2}$},mycircle2,left =0.5cm of a3] (a2) {$a_{2}$};
      \node[label =below:{$t_1 \leq \tilde{t}_{a_1}$},mycircle2,left = 0.5cm of a2] (a1) {$a_{1}$};
      
      \node[label =below:{$\tilde{t}_{N+1}  \leq t_N$},mycircle2,below = of aNp] (aNp2) {$a_{N+1}$};
      \node[label =below:{$\tilde{t}_{N}$},mycircle2,left =5cm of aNp2] (aN2) {$a_{N}$};
      \node[label =below:{$\tilde{t}_{a_3}$},mycircle2,left =1cm of aN2] (a32) {$a_{3}$};
      \node[label =below:{$\tilde{t}_{a_2}$},mycircle2,left =0.25cm of a32] (a22) {$a_{2}$};
      \node[label =below:{$t_1 \leq \tilde{t}_{a_1}$},mycircle2,left = 0.25cm of a22] (a12) {$a_{1}$};

    \foreach \i/\j/\txt/\p in {
      c0/c1//below,
      c0/a1//below,
      c0/a12//below}
       \draw [myarrow] (\i) -- node[sloped,font=\small,\p] {\txt} (\j);

    \foreach \i/\j/\txt/\p in {
      cA/cAp//below,
      a1/a2//below,
      a2/a3//below,
      aN/aNp//below,
      a12/a22//below,
      a22/a32//below,
      aN2/aNp2//below}
       \draw [dashed,->] (\i) -- node[sloped,font=\small,\p] {\txt} (\j);

\node at ($(a3)!.5!(aN)$) {\ldots };
\node at ($(a32)!.5!(aN2)$) {\ldots };
\end{tikzpicture}

\caption{When time is not verifiable, the adversary needs to report that block $a_1$ was created after block $c_0$ and timestamp (and reveal) $a_{N+1}$ on or before $t_N$. In between, the adversary can choose any increasing sequence of timestamps. Because we have assumed deterministic mining and the honest miners have mined $c_0$, we assume that the earlier time possible for $\tilde{t}_{a_1}$ is $t_1$.}
\label{fig:unveriftime}
\end{figure*}

We now consider two extremes of this flexible notion of time. In the first, time is verifiable and the adversarial miner must behave like an honest miner when it comes to reporting time. In the second, time is not verifiable and the adversary is able to report timestamps with a lot more flexibility.

\paragraph{Verifiable timestamps.}
When timestamps are \textbf{verifiable}, the timing of the adversary's actions are observable by the honest miners which essentially forces the adversary to be honest as well. By this, we do not mean that honest miners can see everything the adversary does. We have in mind a situation where it is essentially impossible to falsify time due to the presence of a well-designed accept/reject protocol. It is also possible for a blockchain to feature some hardware technology like Intel SGX which provides trusted time. Figure \ref{fig:veriftime} shows what these timestamps look like when the adversary is reporting honestly.

\paragraph{Unverifiable timestamps.}
When timestamps are \textit{\textbf{unverifiable}}, and the honest miners are naive and do not use any additional safeguards, nothing prevents the adversary from choosing a timestamp for block $a_i$ that is different from the time he found the nonce for block $a_{i-1}$. We allow for any timestamp $\tilde{t}_i$ so long it is after the timestamp $\tilde{t}_{i-1}$. We also assume that blocks cannot contain a timestamp greater than the honest miner's time when the alternative chain is revealed. Therefore, an adversarial and possibly dishonest miner is free to report any strictly increasing sequence of $\tilde{t}_{i}$ subject to those caveats. Of course, actual time taken to mine a block continue to obey Equation (\ref{blockfindtime}) and the difficulty level continues to evolve following Equation (\ref{diffadj}) for the chain the adversary is mining on. To highlight the flexibility that the adversary possesses, Figure \ref{fig:unveriftime} displays two possibilities for chain $a_i$.

\section{Adversary's Optimal strategies} 
\subsection{Verifiable Timestamps}
In this situation, the adversary's optimally chooses $M_i < M_a$ for $i \in \lbrace 1, \dots , N\rbrace$ to construct $a$.  

\paragraph{The naive approach.} 
It is easy to see why choosing $M_i = M_a$ for all $i$ will not work. $a_{1}$ has a difficulty level of 1 and is mined by the adversary in $1/M_a < 1$. Therefore, \[T_{1}=t_A + \frac{1}{M_a} - t_{1}.\] $t_A$ is the time the adversary commences the attack. $1/M_A$ is the time it took for a block with difficulty 1 to be mined, and $t_1$ is the time block $c_0$ was found and hence is the timestamp recorded in block $a_{1}$.
Consequently, \[d_{2} = \frac{d_{1}}{T_{1}}  = \frac{M_a}{(1 + M_a(A-1))}  \]. \[T_{2} = \frac{d_{2}}{M_a} =  \frac{1}{(1 + M_a(A-1))} \] and \[d_{3} = M_a\] In other words, the adversary will gain from applying $M_a > 1$ to a single block with difficulty 1. In addition, the adversary will enjoy the delay of $A - 1$ for one single block and thereafter, extend chain $a$ at the target block growth rate of $1$ just like c. In total, $a_1$ and $a_2$ would have taken the adversary \[\frac{1}{M_a} + \frac{1}{(1 + M_a(A-1))}\] which clearly approaches zero for large $M_a$. However, the difficulty adjustment rule ensures that after two blocks the adversary will end up never catching up with $c$ because both chain will grow at rate 1 per unit time. In a probabilistic setting, the best that the adversary can do is to reach a random walk with an initial deficit of $A - 2 $.


\paragraph{Optimal mining of $k$ blocks.}
We now solve the adversary's problem of optimally allocating mining power if it desires to construct $k$ blocks as quickly as possible. Later, we will relate $k$ to the problem of overtaking the canonical chain $c$. Also, we ignore the fact that there is an added speed bonus from the delay $A - 1$ since this is a one-time bonus that only serves to complicate the math with no additional insights.

The adversary's optimization problem is to
\begin{eqnarray*}
        \min_{M_i} &\sum_{i = 1}^{k} T_i \\
	\text{s.t.} &M_i \leq M_a \\
        &T_i = \frac{d_i}{M_i}\\
        &d_{i+1} = \frac{d_i}{T_i}\\
        &d_1 = 1
\end{eqnarray*}
This can be restated as 
\begin{eqnarray*}
  \min_{M_{i}} &\frac{1}{M_{1}}+\frac{M_{1}}{M_{2}}+...+\frac{M_{k-1}}{M_{k}}\\
  \text{s.t.} & M_{i}\leq M_{a}
\end{eqnarray*}

Clearly $M_{k}=M_{a}$ is optimal. Then by Arithmetic-Geometric mean
inequality, the smallest value is $k M_a^{-\frac{1}{k}}$, which is attained uniquely when all terms in the summation are equal. Hence the solution is 
\begin{equation}
M_{i}=M_{a}^{\frac{i}{k}} \label{veriftimestrat}    
\end{equation}

The adversary initiates the attack with $M_1 = M_a^{1/k}$ and increases the mining capacity by a factor of $M_1 > 1$ for $k$ blocks. Consequently, the adversarial chain adds a block every $1/M_1 < 1$ time units and is able to gain on the canonical chain. What we have left unanswered is the amount of mining capability $M_a$ required to overcome the adversary's deficit of $A$ blocks which we address next.

\paragraph{Overtaking the canonical chain.}
While the adversary is constructing chain $a$, the canonical chain $c$ is still growing at rate 1. Therefore, the adversary must maintain this scaling up of mining capacity to construct as many blocks at it takes to \textit{gain} $A$ blocks on the canonical chain so that the latest block it mines is ahead of $c_N$ by a single block.
\begin{eqnarray}
	k = A + \frac{k}{M_1}  \label{overtake}
\end{eqnarray}  

Equation (\ref{overtake}) relates number of blocks mined to conduct the attack $k$ and the duration of the attack  $\frac{k}{M_1}$ to the initial deficit of $ A $ blocks. The honest miners have a head-start of $A$ and continue to mine $\frac{k}{M_1}$ in the time it takes for the adversary to mine $k$ blocks. Combining Equation (\ref{overtake}) with (\ref{veriftimestrat}) yields
\begin{equation}\label{optverif}
	A = k\left(1 - \frac{1}{\sqrt[k]{M_a}}\right)
\end{equation} 
which links the three fundamental quantities of the initial deficit $A$ to the duration of the optimal attack $k$ and the adversary's mining capacity $M_a$. For a fixed $A$, note that the mining capacity required grows the faster one intends to attack. As $M_a$ approaches infinity, $k$ approaches $A$. Indeed, instantaneously overcoming a deficit of $A$ would require an infinite amount of $M_a$. 

For example with $M_a = 16$, dedicating $M_a = \left\{2,4,8,16\right\}$ would allow for a gain of 4 blocks in 2 units of time since each block takes half a unit of time. Therefore, the adversary can overtake on $A = 2$ if it possessed at least $M_a = 16$ times more mining power than the honest miners. One quickly realizes that the mining power needed to overcome larger values of $A$ blows up the required $M_a$ due to the $k$-th root since $k > A $ for finite $M_a$. 

$M_a$ also decreases in the number of blocks taken $k$. For $A = 2, k=3; M_a = 27$. For $A = 2, k=4; M_a = 16$ and for large $n$, $M_a$ is about 7.5. For $A = 3, k = 4; M_a = 256$. $M_a$ approaches about 20 for large $n$. In other words, to overtake the canonical chain from a deficit of two blocks over an arbitrarily long attack, the adversary would need at least 7.5 times more mining power than honest miners. This corresponds to an adversarial miner who controls about 90pct of the hash rate. To accomplish the same feat in 3 blocks which is the fastest possible, the adversarial miner would need to be in control of about 95pct of the mining power. In short, the difficulty adjustment protocol coupled with a protocol for validating timestamps provides vast amounts of protection against longest-chain attacks particularly those originating from a long range. 

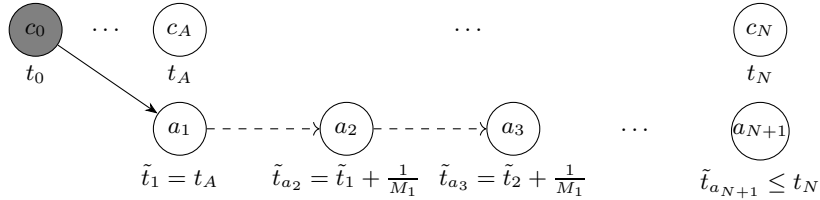
\begin{figure*}[!ht]
\begin{tikzpicture}[
      mycircle/.style={
         circle,
         draw=black,
         fill=gray,
         fill opacity = 0.3,
         text opacity=1,
         inner sep=0pt,
         minimum size=20pt,
         font=\small},
      mycircle2/.style={
         circle,
         draw=black,
         fill=white,
         fill opacity = 0.3,
         text opacity=1,
         inner sep=0pt,
         minimum size=20pt,
         font=\small},
      myarrow/.style={-Stealth},
      myarrow2/.style={-{Implies},double},
      node distance=0.6cm and 1.2cm
      ]
      \node[label =below:{$t_0$},mycircle] at (-5, 0) (c0) {$c_0$};
      \node[label =below:{$t_A$},mycircle2,right=of c0] (cA) {$c_{A}$};
      \node[right= 7cm of cA,label =below:{$t_{N}$},mycircle2] (cN) {$c_N$};
      
      \node[label =below:{$\tilde{t}_{1} = t_A$},mycircle2,below = of cA] (a1) {$a_{1}$};
      \node[label =below:{$\tilde{t}_{a_2} = \tilde{t}_1 + \frac{1}{M_1}$},mycircle2,right = 1.5cm of a1] (a2) {$a_{2}$};
      \node[label =below:{$\tilde{t}_{a_3} = \tilde{t}_2 + \frac{1}{M_1}$},mycircle2,right = 1.5cm of a2] (a3) {$a_{3}$};
      \node[label =below:{$\tilde{t}_{a_{N+1}} \leq t_N$},mycircle2,below = of cN] (aNp) {$a_{N+1}$};
      
    \foreach \i/\j/\txt/\p in {
      c0/a1//below}
       \draw [myarrow] (\i) -- node[sloped,font=\small,\p] {\txt} (\j);

    \foreach \i/\j/\txt/\p in {
      a1/a2//below,
      a2/a3//below}
       \draw [dashed,->] (\i) -- node[sloped,font=\small,\p] {\txt} (\j);

\node at ($(a3)!.5!(aNp)$) {\ldots };
\node at ($(c0)!.5!(cA)$) {\ldots };
\node at ($(cA)!.5!(cN)$) {\ldots };
\end{tikzpicture}
\caption{The adversary initiates the attack with $M_1 = M_a^{1/N}$ and increases it by factor of $M_1$ each block so that $M_i = M_1^{i}$. Therefore, every block takes $1/M_1 < 1$ to mine. We ignore the initial $A-1$ delay bonus since that only helps the adversary, and only for two blocks.}
\label{fig:optverif}
\end{figure*}

We illustrate this optimal strategy in Figure \ref{fig:optverif}.

\subsection{Unverifiable Timestamps} 
While verifiable timestamps are useful, they are not easily attainable without reliance on more frequent communication (e.g., gossip, clock synchronization, etc.) and subjectivity. One may desire a blockchain design with more objectivity, or perhaps, reduce the bandwidth taken up by frequent communication to obtain a network time. We now study such a scenario to shed light on these questions.

While the adversary has flexibility in reporting $\tilde{T}_i$, the following conditions must still be respected.
\begin{eqnarray*}
	\tilde{T}_i &>& 0 \\
	\sum_{i = 1}^{N} \tilde{T}_i &\leq& N - 1\\
	\tilde{t}_{a_1} &\geq& t_1 = 1\\
        \tilde{t}_{N+1} &\leq& t_N = N
\end{eqnarray*}

The first equation is equivalent to an assertion that the time reported in a block must be ahead of the time reported in its parent. Also, a block discovery time of zero implies infinite mining power which is impossible. The second condition simply states that the reported block times must occur in the past at the time of revelation to the honest miners which occurs when the honest miners are on block-height $N$. The third condition states that both chains share a common ancestor at height 0. The final condition which follows from the previous two simply states that the adversary must be a block ahead of the honest miners at the time the alternative chain is revealed.

\paragraph{Optimal reporting of $\tilde{t}_i$.} 
It is obvious that $M_i = M_a$ irrespective of $d_i$. The adversary's optimization problem is to minimize the actual time taken to mine while deploying full mining power by choosing the times to report which influences $d_i$ subject to the conditions above. Leaving the blockheight of overtaking as an unknown parameter, $N$, the adversary's problem is
\begin{eqnarray*}
	\min_{\tilde{T}_i} & \sum_{i = 1}^{N} T_i \\
	\text{s.t.} &	d_{i+1} = \frac{d_{i}}{\tilde{T}_i}\\
	&T_i = \frac{d_i}{M_a} \\
	&\tilde{T}_i > 0 \\
	&\sum_{i = 1}^{N} \tilde{T}_i \leq N - 1\\
	&\tilde{t}_{a_{1}} \geq t_1	   \\
        &d_1 = \frac{d_0}{T_0} = 1 
\end{eqnarray*}

Rewriting, we get
\begin{eqnarray}
	\min_{\tilde{T}_i} &\frac{d_1}{M_a} + \frac{d_1/\tilde{T}_{1}}{M_a} + \frac{d_1/(\tilde{T}_{1}\tilde{T}_{2})}{M_a} + \dots + \frac{d_1/\prod_{i = 1}^{N-1}\tilde{T}_i }{M_a}  \label{unverifprob}\\
    \text{s.t.} &\sum_{i = 1}^{N} \tilde{T}_i \leq N - 1 \nonumber
\end{eqnarray}

Noting that $d_{1} = 1$ since $d_{1} = d_0/T_0 = d_0 = 1$ and that $M_a$ is some constant, we get
\begin{eqnarray*}
	\min_{\tilde{T}_i} & 1 + \frac{1}{\tilde{T}_{1}} + \frac{1}{\tilde{T}_{1} \tilde{T}_{2}} + \dots + \frac{1}{\prod_{i = 1}^{N-1}\tilde{T}_i}  \\
	\text{s.t.} &\sum_{i = 1}^{N-1} \tilde{T}_i + \tilde{T}_N \leq N - 1
\end{eqnarray*}

The objective is to minimize the actual time spent on the attack by optimally reporting fake timestamps which affect the difficulty of the next block. For example, the time that $a_1$ takes is fixed at $1/M_a$ but $a_2$ will depend on how long $a_1$ was reported to have taken. Finally, the difficulty of the last block $a_N$ will depend on all the reported times taken in the previous blocks. Since the objective is to report as quickly as possible, the adversary will claim that the last block was found approximately instantaneously even though it took the adversary $\frac{d_1/\prod_{i = 1}^{N-1}\tilde{T}_i }{M_a}$ time to mine. Likewise, $\tilde{t}_1 = t_1$ because reporting any later time only makes subsequent blocks more difficult to mine. 

As a result of this setup, the adversary will claim that block $N$ was done in zero time (effectively claiming it has $M_a$ arbitrarily large) making the difficulty for block $N+1$ infinitely difficult after the honest miners naively move to chain $a$. This rather ridiculous solution is optimal because our game ends and there is no additional reward from continuing the chain. To avoid this arguably absurd result, we would need to additionally specify what the honest miners believe about $M_a$. For instance, if the honest miners believe that $M_a$ lies in some lower and upper bound, $M_a \in \left[M_l,M_u\right]$, the adversary would have to report timestamps that imply $M_N = M_u$. This doesn't change any  of the key ideas we take away from our analysis but adds an analytical burden. 

Additionally, the constraint must bind since reporting any longer time to find the first block makes all subsequent blocks easier to work on. Omitting the first term gives us.
\begin{eqnarray*}
	\min_{\tilde{T}_i} & \frac{1}{\tilde{T}_1} + \frac{1}{\tilde{T}_1 \tilde{T}_{2}} + \dots + \frac{1}{\prod_{i = 1}^{N-1}\tilde{T}_i}  \\
	\text{s.t.} &\sum_{i = 1}^{N-1} \tilde{T}_i = N-1 = N^* \\
\end{eqnarray*}

Now let's factor out $1/\tilde{T}_1$ and replace $\tilde{T}_1 = N^*- \sum_{i = 2}^{N-1} \tilde{T}_i$ to obtain
\begin{eqnarray}
	\min_{\tilde{T}_2, \dots,  \tilde{T}_{N-1}} \frac{1}{N^* - \sum_{i = 2}^{N-1} \tilde{T}_i} \left(1 + \frac{1}{\tilde{T}_{2}} + \frac{1}{ \tilde{T}_{2}\tilde{T}_{3}} + \dots + \frac{1}{\prod_{i = 2}^{N-1}\tilde{T}_i}  \right) 
    \label{eq:unverifiableTS}
\end{eqnarray}

Appendix \ref{appendix} presents the solution to the minimization problem. 

\section{How important is timestamp verifiability?}
We now compare the adversary's optimal strategy between the two regimes. As we discussed earlier, a longest-chain attack does not make sense for adversaries who possess a small majority of mining power. We consider two adversaries, one with 75pct of the mining capacity and the other with 99pct of the mining capacity. While it may be comical to think about a conventional miner with such capabilities, this risk is a lot more tangible when we allow for quantum computing possibilities \cite{Bard2021QuantumAO}. Such a risk may also be a lot more conceivable for POW blockchains where the overall hash rate is much lower than of Bitcoin's for example.

\subsection{Results}
Table \ref{table:results} reports what values of initial deficits $A$ an adversary can overcome if it mounts an attack where it selfishly mines $N = 3,5,10, 20, 100$ blocks. We also report the time taken $T^*(N) = t_{a_{N+1}} - t_{a_1}$ to mine the alternative chain. As the adversary mines $N$ blocks in such a fashion, the honest miners would have extended chain $c_A$ by $T^*(N)$ blocks. Therefore, the largest $A$ the adversary could have overcome would be given by $N - T^*(N)$. Integer constraints are ignored.

\begin{table*}[t]
\begin{center}
\begin{tabular}{ c V{2.5} c  c V{2.5} c c  V{4} c c V{2.5} c c }
\hline
\hline
& \multicolumn{4}{c V{4}}{\textbf{$M_a = 3$ (75pct of total capacity)}} & \multicolumn{4}{c}{\textbf{$M_a = 99$ (99pct of total capacity)}} \\
\cline{2-9}
\textbf{N} & \multicolumn{2}{c V{2.5}}{\textbf{Verifiable Time}} & \multicolumn{2}{ c V{4}}{\textbf{Unverifiable Time}} & \multicolumn{2}{ c V{2.5}}{\textbf{Verifiable Time}} & \multicolumn{2}{ c }{\textbf{Unverifiable Time}} \\ 
 & $T^*(N)$ & $A_{max}$ & $T^*(N)$ & $A_{max}$ & $T^*(N)$ & $A_{max}$ & $T^*(N)$ & $A_{max}$ \\
\hline
\hline
\textbf{3} & 2.08 & 0.92 & 0.96 & 2.04 & 0.65 & 2.35 & 0.03 & 2.97 \\
\textbf{5} & 4.01 & 0.99 & 1.43 & 3.57 & 1.99 & 3.01 & 0.04 & 4.96 \\
\textbf{10} & 8.96 & 1.04 & 2.21 & 7.79 & 6.32 & 3.68 & 0.07 & 9.93 \\
\textbf{20} & 18.93 & 1.07 & 3.04 & 16.96 & 15.89 & 4.11 & 0.09 & 19.91 \\
\textbf{100} & 98.91 & 1.09 & 4.37 & 95.63 & 95.51 & 4.49 & 0.13 & 99.87 \\ \hline

\end{tabular}
\end{center}
\caption{Time taken to mount an attack of $N$ blocks and the corresponding maximum lead $A$ the adversary can overcome for $M_a = 3$ and $M_a = 99$.}
\label{table:results}
\end{table*}

\subsection{Discussion: Practical Implications}
What we call a block should not be taken literally. In reality, our block represents an epoch. Secondly, our assumptions on the extremes of timestamp verifiability also should not be taken literally. What matters is how flexible timestamps can be relative to epoch length. For instance, Bitcoin's timestamps can be any time in a 3-hour window and be accepted. Its epoch length is 2016 blocks which will take about 2 weeks to mine. As a ratio, the relative flexibility approaches zero. This means that Bitcoin is probably very close to our setting with verifiable time and it suggests that an adversary controlling 75pct of the mining capacity will be able to start an entire epoch behind the canonical chain and overtake it after 4 epochs have elapsed on the canonical chain. Monero and Bitcoin Cash, recalculate the block adjustment every block using the previous day's worth of blocks. The degree of time reporting flexibility is similar to Bitcoin's. As a ratio, these two blockchains would be further away from perfect time verifiability compared to Bitcoin. We do not attempt to extrapolate the effects of the moving average or comment on the various approaches in determining which reported timestamp is valid. Our analysis is deliberately kept very simple so that the forces at work are transparently characterized. 

It is also important to note that our results are silent on how vulnerable blockchains are to the best possible strategy an adversary can mount. We are only commenting on this particular strategy which is the selfish mining of an alternative longest chain.

\paragraph{Takeaway 1: Verifiable timestamps diminish the efficacy of $M_a$.}
Observe that with both cases for $M_a$, the time it takes to construct more blocks increases with the number of blocks constructed. Since the canonical chain is growing at the same time, the adversary will need to control huge amounts of mining power in order to overcome small leads. In other words, an adversary will find it very difficult to start an alternative chain that is more than a few blocks behind the leading block and overtake it. The crucial insight here is that the \textit{best} strategy the adversary can employ is to scale up its mining efforts following a power law and power law progressions ramp up very quickly. It also sets limits on how far ahead the target chain can be for an adversary with capacity $M_a$. For example, an adversary with $M_a = 3$ cannot attempt the longest-chain attack starting from 2 blocks behind the canonical chain. It will \textbf{never} catch up no matter how long it mines. 

\paragraph{Takeaway 2: Unverifiable timestamps lead to approximately linear attack duration.}
With unverifiable timestamps, the time taken to construct $N$ blocks is approximately linear in $N$ under our assumption that $\tilde{t}_{a_1} = t_{c_1}$. This implies that an adversary possessing mining power greater than 51pct can and will catch up any distance A provided it continues selfishly mining for long enough. 

\paragraph{Future Work.}
We solved for the optimal attack an adversary can mount against naive honest miners assuming a very limited action set for the adversary. For instance, the adversary is not allowed to mine on the main chain. If that action was allowed, it is very easy to show that the adversary can improve on its desired outcome by mining on the main chain whenever its difficulty is low, and, leaving it to mine on its own chain whenever the difficulty increases. This is commonly known as chain hopping. Quantifying this optimal action depending on time verifiability is an immediate extension of this paper. 

\bibliographystyle{abbrv}
\bibliography{bibliography}

\appendix

\noindent {\LARGE \textbf{Appendix}}

\section{Solving Equation \ref{eq:unverifiableTS}}
\label{appendix}

Clearly, each choice variable is bounded away from zero as otherwise, the objective would diverge. So there is some $\epsilon>0$ such that $\tilde{T}^*_i > \epsilon$. Consider the same minimization problem with constraints $\tilde{T}_i \geq \epsilon$. Then we know the solution to the new problem is interior. Then the solution is given by First Order Conditions (FOCs). But the FOCs of the new problem are the same as the FOCs of the original problem. If the FOCs yield a unique solution, it is the unique minimizer of the original problem. Next, we take the FOCs of the original problem.

So, zero gradient for any $\tilde{T}_x$
\begin{eqnarray*}
    \frac{1}{\left(N^* - \sum_{i = 2}^{N-1} \tilde{T}_i\right)^2} \left(1 + \frac{1}{\tilde{T}_{2}} + \frac{1}{ \tilde{T}_{2}\tilde{T}_{3}} + \dots + \frac{1}{\prod_{i = 2}^{N-1}\tilde{T}_i}  \right) = \\\frac{1}{\left(N^* - \sum_{i = 2}^{N-1} \tilde{T}_i\right)}\left(\frac{1}{\tilde{T}_x}\sum_{k=x}^{N-1}\frac{1}{\prod_{j=2}^{k}\tilde{T}_{j}}\right)
\end{eqnarray*}
gives
\begin{eqnarray*}
    \frac{1}{\left(N^* - \sum_{i = 2}^{N-1} \tilde{T}_i\right)} \left(1 + \frac{1}{\tilde{T}_{2}} + \frac{1}{ \tilde{T}_{2}\tilde{T}_{3}} + \dots + \frac{1}{\prod_{i = 2}^{N-1}\tilde{T}_i}  \right) =\\ \frac{1}{\tilde{T}_x}\sum_{k=x}^{N-1}\frac{1}{\prod_{j=2}^{k}\tilde{T}_{j}}
\end{eqnarray*}
Note that the LHS is the objective which is a constant when evaluated at its optimal. Immediately, note as well that the higher $x$ is, the fewer terms there are in the sum on the RHS. A lower $x$ also contains all the terms in summation from $x+1$ plus one more. Hence, $T_x$ has to be decreasing in $x$ which is of course, again, consistent with the idea that a smaller $T_x$ increases the difficulty on the next block and must be avoided. 

Next, note that we can break up the sum so that:
\begin{equation*}
\begin{gathered}
    \frac{1}{\tilde{T}_x}\sum_{k=x}^{N-1}\frac{1}{\prod_{j=2}^{k}\tilde{T}_{j}} = \frac{1}{\tilde{T}_{x+1}}\sum_{k={x+1}}^{N-1}\frac{1}{\prod_{j=2}^{k}\tilde{T}_{j}} \\
    \frac{1}{\tilde{T}_x}\left(\sum_{k=x}^{x}\frac{1}{\prod_{j=2}^{k}\tilde{T}_{j}} + \sum_{k=x+1}^{N-1}\frac{1}{\prod_{j=2}^{k}\tilde{T}_{j}}\right) = \frac{1}{\tilde{T}_{x+1}}\sum_{k={x+1}}^{N-1}\frac{1}{\prod_{j=2}^{k}\tilde{T}_{j}}\\
    \frac{\frac{1}{\prod_{j=2}^{x}\tilde{T}_{j}}}{\sum_{k=x+1}^{N-1}\frac{1}{\prod_{j=2}^{k}\tilde{T}_{j}}}  = \frac{\tilde{T}_x}{\tilde{T}_{x+1}} - 1
\end{gathered}
\end{equation*}

And then we can multiply the top and bottom of the LHS by $\prod_{j=2}^{x}\tilde{T}_{j}$ to get
\begin{equation*}
\begin{gathered}
    \frac{1}{\sum_{k=x+1}^{N-1}\frac{1}{\prod_{j=x+1}^{k}\tilde{T}_{j}}}  = \frac{\tilde{T}_x}{\tilde{T}_{x+1}} - 1 \\
    \sum_{k=x+1}^{N-1}\frac{1}{\prod_{j=x+1}^{k}\tilde{T}_{j}}  = \frac{\tilde{T}_{x+1}}{\tilde{T}_{x} - \tilde{T}_{x+1}}  
\end{gathered}
\end{equation*}

Some manipulation yields
\begin{equation}\label{optrep}
\begin{gathered}
    \tilde{T}_{x}  = \frac{\tilde{T}_{x-1} - \tilde{T}_{x}}{\tilde{T}_{x} - \tilde{T}_{x+1}}  \text{    for    } x = 2...N-2\\
    \frac{\tilde{T}_{N-1}}{\tilde{T}_{N-2}-\tilde{T}_{N-1}} = \frac{1}{\tilde{T}_{N-1}}\\
    \sum_{i = 1}^{N-1} \tilde{T}_i = N^*
\end{gathered}
\end{equation}
which fully characterizes the solution.

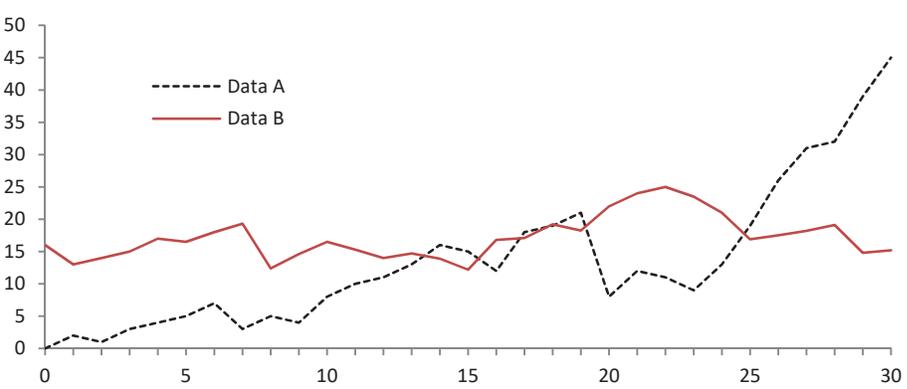
\end{document}